# Failure Mechanism of True 2D Granular Flows

Cuong T. NGUYEN[1,3], Ha H. BUI[2], and Ryoichi FUKAGAWA[3]

[1]*Institute of Mechanics, Vietnam Academy of Science and Technology, 18 Hoang Quoc Viet, Cau Giay, Hanoi, Vietnam*

[2]*Department of Civil Engineering, Monash University, Clayton Campus, Melbourne, VIC 3800, Australia*

[3]*Department of Civil Engineering, Ritsumeikan University, 1-1-1 Nojihigashi, Kusatsu, Shiga 525-8577, Japan*

**Keywords:** Granular Flow, 2D Granular Column Collapse, Run-out Distance, Failure Mechanism

Most previous experimental investigations of two-dimensional (2D) granular column collapses have been conducted using three-dimensional (3D) granular materials in narrow horizontal channels (i.e., quasi-2D condition). Our recent research on 2D granular column collapses by using 2D granular materials (i.e., aluminum rods) has revealed results that differ markedly from those reported in the literature. We assume a 2D column with an initial height of $h_0$ and initial width of $d_0$, $a$ defined as their ratio ($a = h_0/d_0$), a final height of $h_\infty$, and maximum run-out distance of $d_\infty$. The experimental data suggest that for the low $a$ regime ($a \leq 0.65$) the ratio of the final height to initial height is 1. However, for the high $a$ regime ($a \geq 0.65$), the ratio of $a$ to $(d_\infty - d_0)/d_0$, $h_0/h_\infty$, or $d_\infty/d_0$ is expressed by power-law relations. In particular, the following power-function ratios ($h_0/h_\infty \approx 1.42\, a^{2/3}$ and $d_\infty/d_0 \sim 4.30 a^{0.72}$) are proposed for every $a \geq 0.65$. In contrast, the ratio $(d_\infty - d_0)/d_0 \approx 3.25 a^{0.96}$ only holds for $0.65 \leq a \leq 1.5$, whereas the ratio $(d_\infty - d_0)/d_0 \approx 3.80 a^{0.73}$ holds for $a \geq 1.5$. In addition, the influence of ground contact surfaces (hard or soft beds) on the final run-out distance and destruction zone of the granular column under true 2D conditions is investigated.

## Introduction

The flow of granular material is commonly observed in engineering applications such as the transport of minerals, powder, or cereals and during geophysical events such as landslides and debris flow. Understanding the mechanisms of granular flow will help to optimize industrial processes and to minimize damage caused by natural disasters. Accordingly, many scientists have been interested in studying this problem both experimentally and by using numerical simulations under two-dimensional (2D) and three-dimensional (3D) conditions.

In the past, most 2D granular flow experiments have been conducted in narrow horizontal flow channels by using 3D materials such as sandy soils or plastic/glass beads. Typical examples of this type of quasi-2D experiment include those of Balmforth and Kerswell (2005), Lube *et al.* (2005), Lajeunesse *et al.* (2005), and Trepanier *et al.* (2010). By using grit, fine glass, coarse glass, and polystyrene as granular materials, Balmforth and Kerswell (2005) reported the following relationship between the final run-out distance and the initial aspect ratio of the granular column: $(d_\infty - d_0)/d_0 \simeq a^{0.9 \pm 0.1}$ for wide channels, and $(d_\infty - d_0)/d_0 \simeq a^{0.65 \pm 0.05}$ or $d_\infty/d_0 \simeq a^{0.55 \pm 0.05}$ for narrow slots, where $a = h_0/d_0$; $h_0$ and $d_0$ are the initial height and width of the column, respectively; $h_\infty$ and $d_\infty$ are the maximum final height and width of the column, respectively. In similar experiments using fine quartz sand, coarse quartz sand, sugar, and rice as granular materials, Lube *et al.* (2005) concluded that $(d_\infty - d_0)/d_0 = 1.2a$ for $a < 1.8$ and $(d_\infty - d_0)/d_0 = 1.9a^{2/3}$ for $a > 2.8$. There was no abrupt break in the curve $(d_\infty - d_0)/d_0$, within the small transitional region of $1.8 < a < 2.8$ between the linear and power-law ranges. Lajeunesse *et al.* (2005), through a series of granular flow experiments using glass beads of diameter $d = 1.15$ mm and $d = 3$ mm as the granular materials, suggested that $(d_\infty - d_0)/d_0 \simeq a$ and $(d_\infty - d_0)/d_0 \simeq a^{2/3}$ for $a \leq 3$ and $a > 3$, respectively. Trepanier and Franklin (2010) repeated the experiments by Lube *et al.* (2005) and Lajeunesse *et al.* (2005) but with randomly arranged granular rods. They reported that the ratio $(d_\infty - d_0)/d_0$ was given by $a^{1.2 \pm 0.1}$ and $a^{0.6 \pm 0.1}$ when $a < (1.1 \pm 0.3)$ and $> (1.1 \pm 0.3)$, respectively.

In addition to the quasi-2D experiments, the mechanism of granular flow is often investigated under axisymmetric conditions (i.e., 3D conditions). Typical experiments of this type include those by Lube *et al.* (2004), Lajeunesse *et al.* (2004), Lajeunesse *et al.* (2005), and most recently Warnett *et al.* (2014). Many authors have used the experimental results described above to verify their 2D/3D numerical models, which they then used to study granular flow scenarios that are difficult to model experimentally. Staron and Hinch (2005), Bui (2007), Bui *et al.* (2006, 2008a, 2008b, 2009), and Trepanier and Franklin (2010) are but a number of authors who have adopted this approach.

As the literature review above shows, most previous 2D granular flow experiments were conducted under quasi-2D conditions. Although such experimental data could be used to validate 2D numerical models, they do not reflect true 2D conditions in the simulations, in plane strain or plane stress conditions. To overcome this knowledge gap, this paper presents the results of a series of column collapse experiments in which aluminum rods are used as the granular materials, i.e., a true 2D condition. To our best knowledge, a true 2D granular column collapse experiment has not yet been conducted in the literature. Herein, we focused on factors that affect the final run-out distance or destruction zone of the granular column, including the characteristics of experimental materials and type of ground contact surface (hard or soft). Notably, we also examined the effect of soil ground quality on the run-out distance and destruction zone of granular columns.

## 1. Experiments

### 1.1 Experimental setup

**Figure 1** shows a schematic diagram of the initial setup for the true 2D granular column collapse experiments, using aluminum rods as soil models. The full design of the experimental model setup is shown in **Figure 1(a)**. However, owing to the symmetrical properties of granular column collapse, only half of the experimental model was considered in the real experiment (**Figure 1(b)**), with the vertical axis (OY) being replaced by a solid wall. The original height and width of the 2D soil layer are $h_0$ and $d_0$, respectively. These parameters, however, were changed during the experiment to investigate their effects on the final run-out distance and destruction zone of the granular column.

### 1.2 Material

Aluminum rods 5 cm in length and with diameters of 1.6 mm and 3.0 mm, mixed at a ratio of 3:2 in weight, were used as the model ground to simulate the true 2D granular flow experiments (Nakai, 2012). The total unit weight of the model ground after construction is 20.4 kN/m$^3$. The soil shear strength parameters of the model ground, including elastic modulus, friction angle, and cohesion, are obtained by conducting direct shear tests or biaxial tests on the aluminum rods. These testing results have been reported by the authors in Bui *et al.* (2008a, 2008b, 2014) and summarized in **Table 1**. Prior to the experiment, the aluminum bars were thoroughly cleaned and dried before being mixed (Nguyen *et al.*, 2013). This was to ensure that the moisture of the model ground was at 0%.

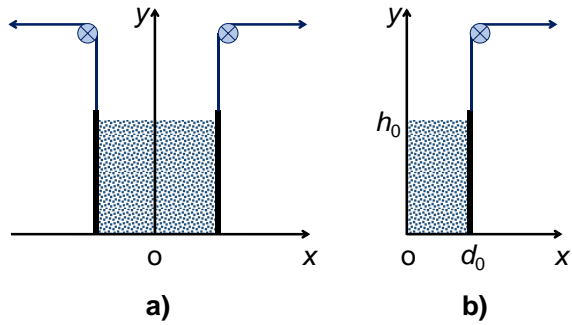

**Fig. 1** Schematic diagram of the experimental model

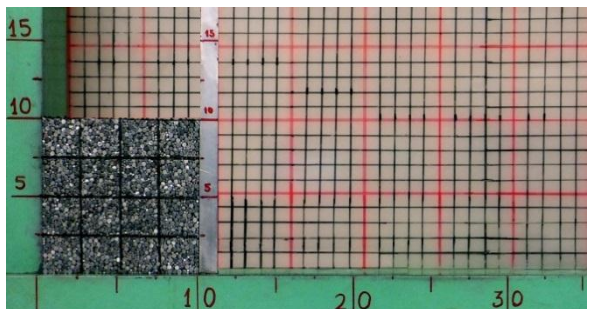

**Fig. 2** Initial setup for the granular collapse experiment with $h_0$ = 100 mm and $d_0$ = 100 mm

**Table 1** Properties of the 2D soil model

| Name | Value | Unit |
| --- | --- | --- |
| Density ($\rho$) | 20.4 | kN/m$^3$ |
| Friction angle ($\phi$) | 21.9 | o |
| Young's modulus (E) | 5.84 | MPa |
| Poisson's ratio ($\nu$) | 0.3 | – |
| Dilatant angle ($\psi$) | 5–7 | o |
| Cohesion (c) | 0 | kPa |

### 1.3 Experimental procedure

A series of experiments were performed at different initial column heights ($h_0$) of 50, 100, 150, and 200 mm. For each of the initial heights, the following granular column widths ($d_0$) were considered: 25, 50, 75, 100, 125, 150, 175, and 200 mm. Accordingly, 48 experiments were conducted in total. Each of the experiments was repeated at least twice, with some repeated up to six times.

**Figure 2** shows the initial setup for the granular column collapse experiments with an initial height and width of 100 mm and 100 mm, respectively. In the experiment, the granular column was constructed by successively placing the ground model in layers of 2.5 cm until the desired initial height was reached. To visualize the failure pattern of the ground model, square grids of 2.5 × 2.5 cm were drawn directly on the soil specimens. In addition, a gridded board was also attached to the steel frame behind the soil sample to allow the visualization of the progressive failure of the granular flow. There was no direct contact between the gridded board and granular rods, neither during nor after the experiment. Furthermore, the front side was open to enable clear visualization of the granular flow.

In each experiment the following steps were repeated: 1) Ensure the model ground is clean and dry by washing the soil sample with alcoholic solutions and then drying it after each experiment; 2) Construct the granular column according to the required experiment dimensions (i.e., initial height and width) and draw square grids (2.5 × 2.5 cm) on the soil specimens; 3) Quickly remove the right wall to allow granular soil to freely move and collapse (care must be taken to ensure that the aluminum rods do not collide with the wall during the collapse process); 4) Record the failure process of the granular soil using a high speed camera and measure the final run-out distance and final failure pattern. The high-speed camera was a Photron type camera, capable of recording 500 frames/s at a resolution of 1024×512 pixels.

## 2. Experimental Observation

Two major failure mechanisms were observed from the serial tests on the hard ground contact surface (**Figure 3**). The failure mechanism depended on the initial ratio of the initial height ($h_0$) to the initial width ($d_0$); at an initial ratio $h_0 / d_0 > 0.65$, the granular column collapsed and formed a conical shape on the top surface (**Figure 3(a)**). On the other hand, when the initial ratio $h_0 / d_0 \leq 0.65$, there was an undisturbed zone on the top surface of the granular column (**Figure 3(b)**). For tests on the soft ground contact surface (made of the aluminum rods) the same failure mechanisms were observed. However, the final run-out distance was slightly different from that on the hard ground surface. Details of the experimental results are summarized below.

## 2.1 Results on hard ground contact surface

For these experiments, we focused on the major failure mechanism of the 2D granular column to determine the relationship between the initial soil height and final run-out distance of the 2D granular column after the collapse.

**Figures 4** and **5** show the progressive collapse of the 2D granular column at several representative time points for the two typical failure mechanisms corresponding to those described in **Figure 3**. The failure mechanism observed in **Figure 4** corresponds to that of **Figure 3(a)**, with a final conical shape on the top surface, whereas that in **Figure 5** corresponds to the mechanism in **Figure 3(b)** with an undisturbed zone. The failure surface, which separates the failure zone from the undisturbed zone, is highlighted in these figures by a dashed line. Comparing the results, the failure surfaces observed with an initial ratio $h_0 / d_0 \geq 0.65$ (**Figure 4**) remain almost unchanged in shape (i.e., a straight line) from those observed with $h_0 / d_0 < 0.65$ (**Figure 5**). The initial soil height of the former case is markedly that of the latter case, mainly owing to the difference in the initial soil volume.

As for the final run-out distance, the experimental results (**Figures 6–8**) show that for $a \leq 0.65$ ($a = h_0 / d_0$) the initial and final heights of the soil column were identical (i.e., $h_0 = h_\infty$). This result is consistent with the quasi-2D experimental data reported by Balmforth Kerswell (2005) and Staron and Hinch (2005). In contrast, for $a > 0.65$, the relationships between the coefficient $a$ and the ratio $(d_\infty - d_0)/d_0$ and between the ratios $h_0/h_\infty$ and $d_\infty/d_0$ were exponential in form.

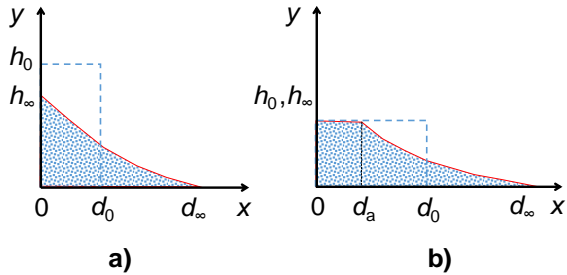

**Fig. 3** Typical failure mechanism of granular column collapse obtained from experiments

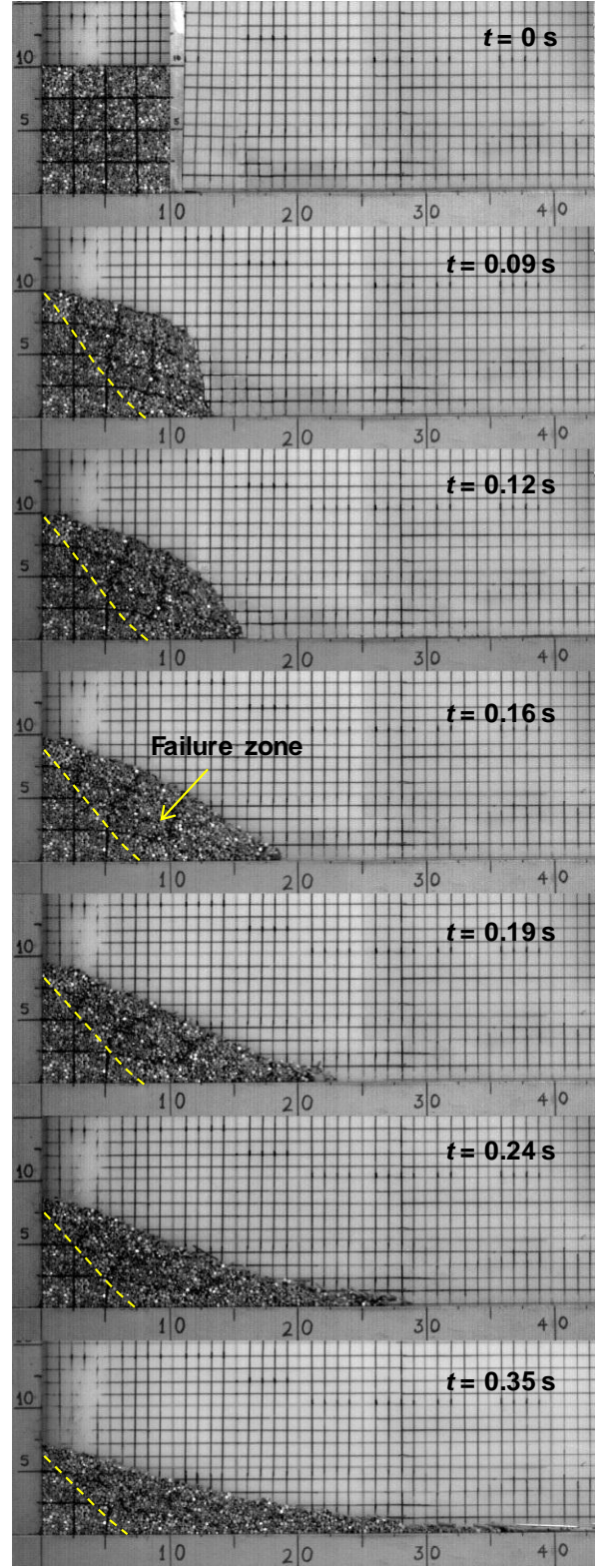

**Fig. 4** The collapse of the granular column with $h_0 = 100$ mm and $d_0 = 100$ mm observed at several time points with a high-speed camera

As shown in **Figure 6**, the exponential relationship between $a$ and the ratio $(d_\infty - d_0)/d_0$ changes at $a = 1.5$, as shown in Eq. (1) below:

$$\frac{d_\infty - d_0}{d_0} \approx \begin{cases} 3.25 a^{0.96} & a < 1.5 \\ 3.80 a^{0.73} & a \geq 1.5 \end{cases} \quad (1)$$

The relationship between $h_0/h_\infty$ and $a$ (**Figure 7**) is approximated by using the following exponential equation:

$$\frac{h_0}{h_\infty} \approx 1.42 a^{2/3} \quad a \geq 0.65 \tag{2}$$

If the experimental data are separated into two data series, consistent with the expression in Eq. (1), the following exponential equations describe the relation between $h_0/h_\infty$ and $a$:

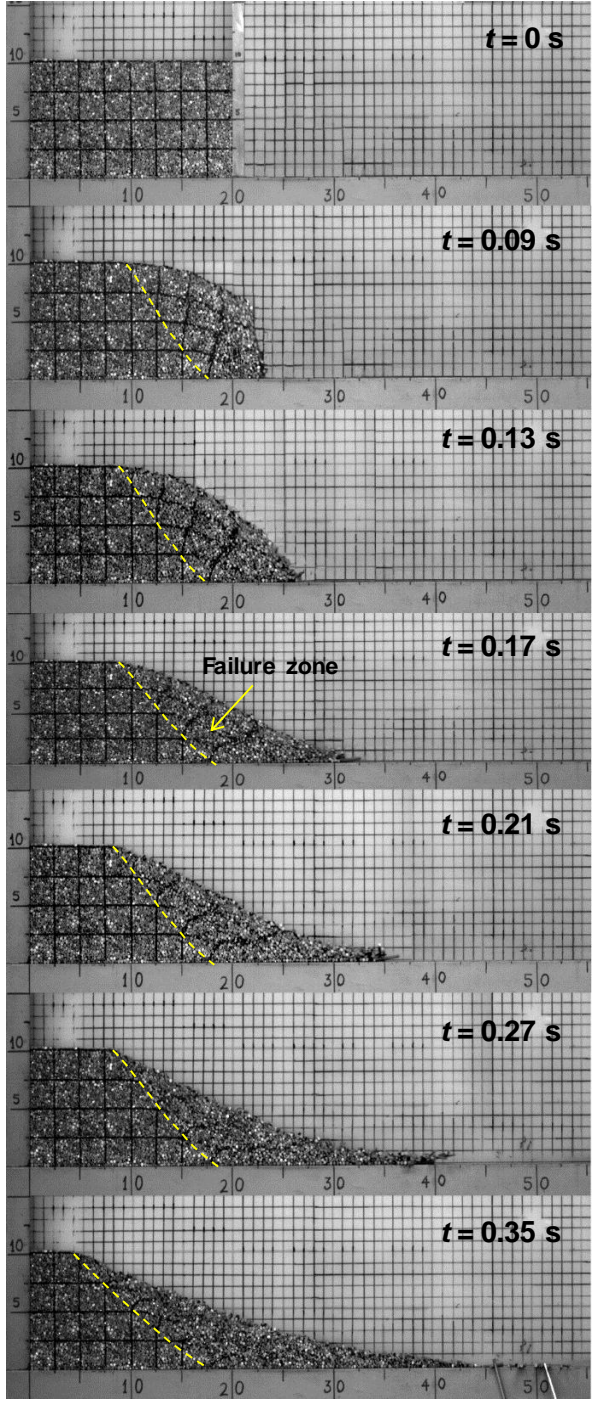

**Fig. 5** The failure process of the granular column ($h_0$ = 100 mm and $d_0$ = 200 mm) at several time points obtained by high-speed camera.

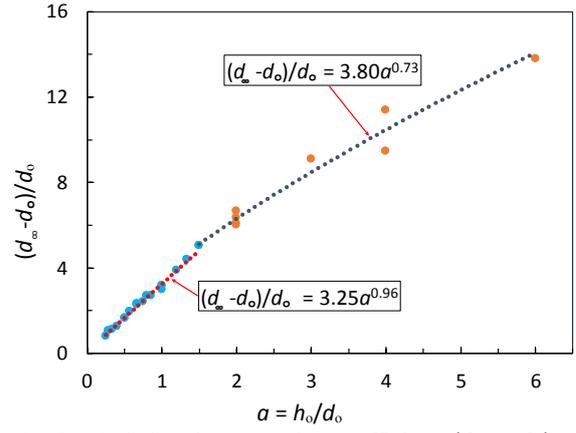

**Fig. 6** Relation between the coefficient $(d_\infty - d_0)/d_0$ and the coefficient $a$ compiled from the experiment results

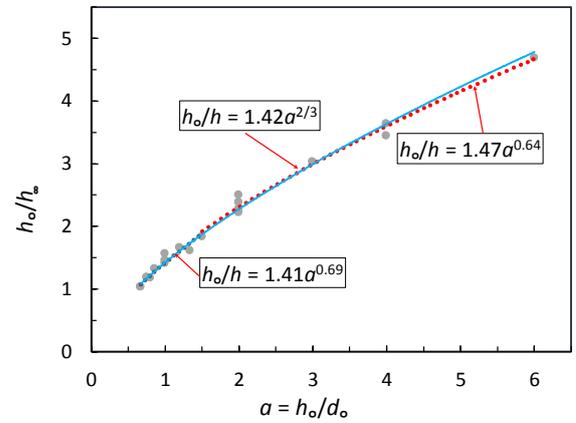

**Fig. 7** Relation between the coefficient $h_0/h_\infty$ and the coefficient $a$ compiled from the experiment results

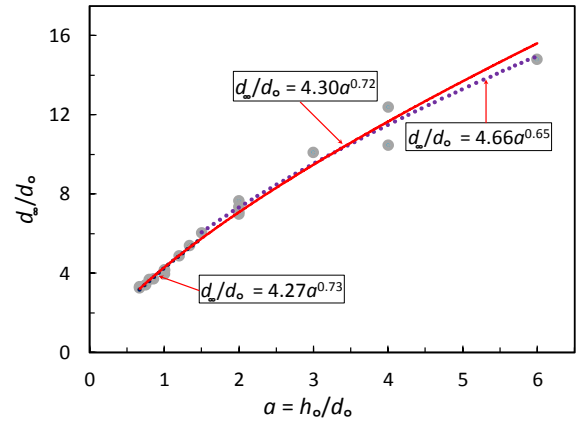

**Fig. 8** Relation between the coefficient $d_\infty/d_0$ and the coefficient $a$ compiled from the experimental results

$$\frac{h_0}{h_\infty} \approx \begin{cases} 1.41 a^{0.69} & a < 1.5 \\ 1.47 a^{0.64} & a \geq 1.5 \end{cases} \tag{3}$$

Similarly, for the ratio $d_\infty/d_0$ (**Figure 8**), the experimental results also show an overall exponential relationship between this ratio and coefficient $a$, which is:

$$\frac{d_0}{d_\infty} \approx 4.30a^{0.72} \quad a \geq 0.65 \tag{4}$$

If the data are again separated into two series with $a = 1.5$ as the breaking point, the following expressions describe the relationship between the ratio $d_\infty/d_0$ and coefficient $a$:

$$\frac{d_\infty}{d_0} \approx \begin{cases} 4.27a^{0.73} & a < 1.5 \\ 4.66a^{0.65} & a \geq 1.5 \end{cases} \tag{5}$$

## 2.2 Results on soft ground contact surface

To investigate the effect of the ground contact surface on the failure mechanism, 2D granular column collapse experiments were conducted on a soft contact surface. These results were compared with those obtained from the experiments conducted on a hard contact surface.

The initial geometric settings and boundary conditions for the current experiment are shown in **Figures 9** and **10**. We only considered the case with a rectangular granular column of $h_0 = 100$ mm and $d_0 = 200$ mm. For the experiment on the soft ground surface, the rectangular granular column was placed on a soft ground layer 2.5 mm in height made of the same type of material as the granular column (i.e., aluminum rods), (**Figure 10**). Each experiment was repeated at least six times to confirm the failure mechanism.

**Figure 11** shows the collapsing process of the granular column on the soft contact surface at several representative time points. Similar to the results on the hard contact surface, granular soils progressed toward the right after removing the retaining wall. The failure surface was almost a straight line from the ground free-surface to the boundary between the initial rectangular granular column (100 × 200 mm) and the soft bed layer (i.e., 2.5 mm thickness). Beyond this boundary, the failure surface was bent and formed a curved surface inside the granular bed layer. This is similar to what is normally observed in a deep-seated landslide failure.

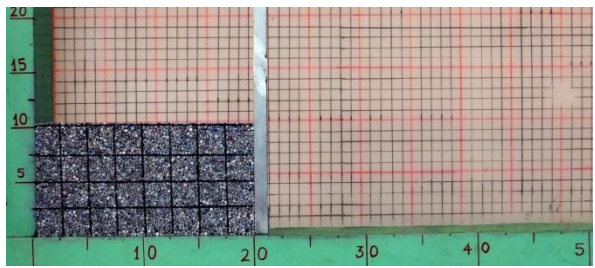

**Fig. 9** Initial experimental setting for hard ground

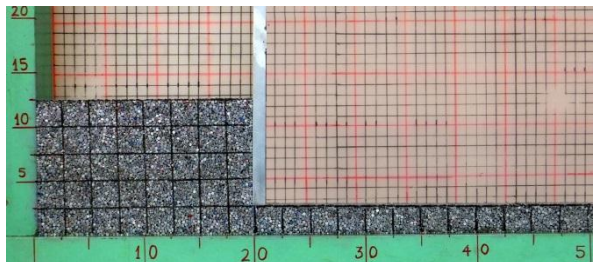

**Fig. 10** Initial experimental setting for soft ground

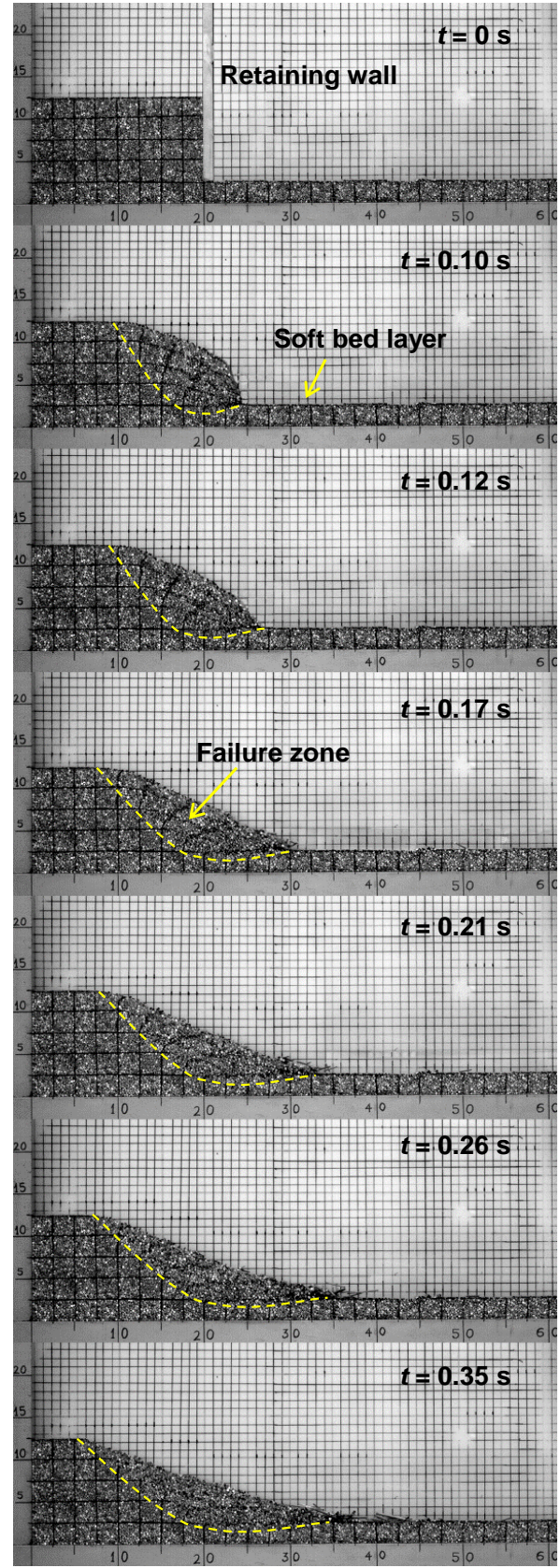

**Fig. 11** The collapsing process of the granular column on soft ground contact surface

The final configurations of the failure surface and ground free-surface observed in both experiments are replotted in **Figures 12** and **13**. Comparing the results, the failure surface observed on the soft ground contact surface differs from that observed on the hard contact surface. For the soft contact surface, part of the kinetic energy from the granular column collapse is transferred to and causes destruction of the soft bed layer during the collapsing process. In contrast, for the hard contact surface, the transfer of kinetic energy to the ground surface is negligible. This explanation is further supported by comparing the final run-out distance of the two cases. As shown in **Figures 12** and **13**, the final run-out distance of the granular column on the hard ground contact surface was approximately 475 mm, whereas that on the soft ground contact surface was approximately 450 mm. This difference is presumably due to the dissipation of the kinetic energy (from the granular column) into the soft ground surface below. Therefore, we conclude that the ground contact surface plays an important role in the final run-out distance of granular flow.

## 3. Discussion

Several research groups have experimentally investigated the failure mechanism of 2D granular flow. Notable studies include those by Balmforth and Kerswell (2005), Lube *et al.* (2005), Lajeunesse *et al.* (2005), and Trepanier and Franklin (2010). However, in all of these studies, the authors adopted quasi-2D models to perform their experiments. In such models, a narrow horizontal channel is typically used with 3D granular materials such as sandy soil, rice, or glass beads to simulate 2D granular flows. Because of the nature of the model and variety of materials used, the reported experimental results for the final run-out distance varied widely across experiments, as summarized below.

In a series of experiments conducted by Balmforth and Kerswell (2005), the authors used grit, fine glass, coarse glass, and polystyrene as the model ground to investigate the failure mechanism of 2D granular flow in a narrow horizontal flow channel with different widths. They found that the final run-out distance of the 2D granular column depends on the width of the flow channel. In particular, they reported the following exponential relationships between the ratio $(d_\infty - d_0)/d_0$ and the coefficient $a$:

$$\frac{d_\infty - d_0}{d_0} = \begin{cases} \lambda a^{0.9 \pm 0.1} & \text{for wide slots} \\ \lambda a^{0.65 \pm 0.05} & \text{for narrow slots} \end{cases} \quad (6)$$

where $\lambda$ is a coefficient depending on the internal friction angle of the used materials and the friction coefficient of the bed contact surface.

Lube *et al.* (2005) also reported experimental results of 2D granular flow by using a quasi-2D model on fine quartz sand, coarse quartz sand, sugar, and rice. They found the following relationship between $(d_\infty - d_0)/d_0$ and $a$:

$$\frac{d_\infty - d_0}{d_0} \approx \begin{cases} 1.2 a & a < 1.8 \\ 1.9 a^{2/3} & a > 2.8 \end{cases} \quad (7)$$

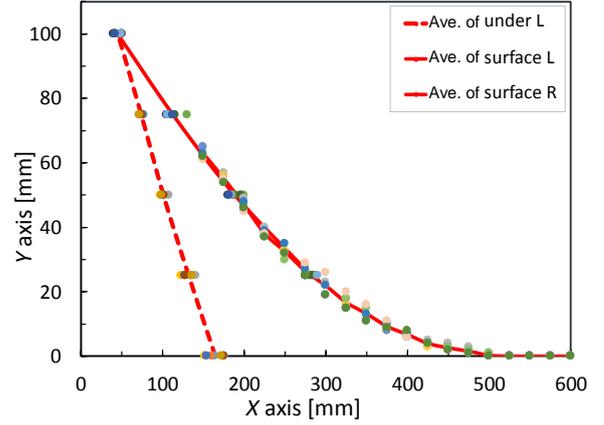

**Fig. 12** The final shapes of failure surface and ground free-surface of the granular collapse on the hard contact surface

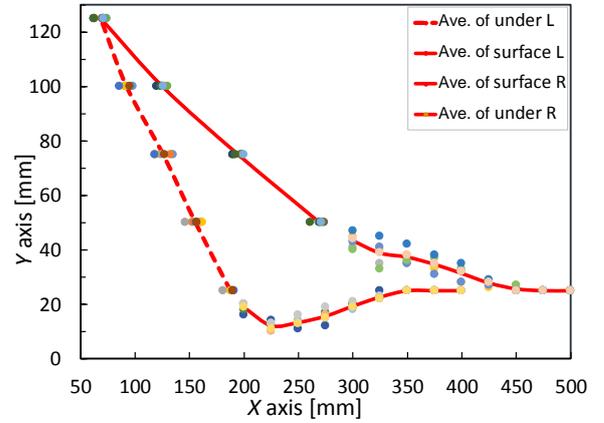

**Fig. 13** The shapes of failure surface and ground free-surface of the granular collapse on the soft contact surface

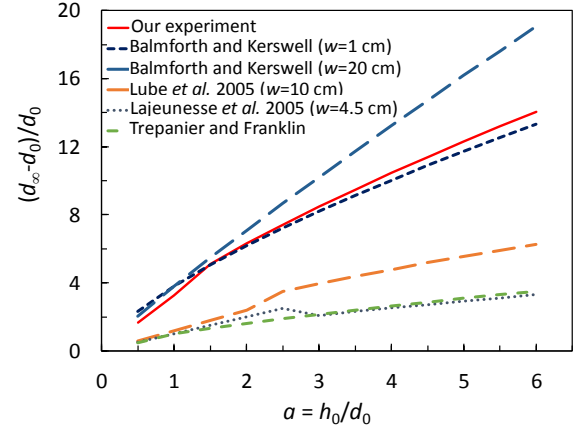

**Fig. 14** The experimental results of quasi-2D granular flow reported by previous authors and results of our true 2D granular flow experiment

In contrast to the result reported by Balmforth and Kerswell (2005), a transition from a linear to exponential relation occurs in the experiments reported by Lube *et al.* (2005).

In another series of experiments that investigated the granular column collapse mechanism, conducted by Lajeunesse *et al.* (2005) on glass beads of diameter $d$ = 1.15 mm and $d$ = 3 mm, a narrow horizontal flow channel of 45 mm in width was used. For this series of experiments, the following relationship between $(d_\infty - d_0)/d_0$ and $a$ was found:

$$\frac{d_\infty - d_0}{d_0} \approx \begin{cases} a & a < 3 \\ a^{2/3} & a > 3 \end{cases} \quad (8)$$

The result is markedly different from that reported by Lube *et al.* (2005), although the transition from a linear to exponential relation is also noted. Most recently, Trepanier and Franklin (2010) used granular rods to investigate the failure mechanism of granular columns under 3D conditions. They reported the following relationship for the final run-out distance:

$$\frac{d_\infty - d_0}{d_0} \approx \begin{cases} a^{1.2 \pm 0.1} & a < 1.1 \pm 0.3 \\ a^{0.6 \pm 0.1} & a > 1.1 \pm 0.3 \end{cases} \quad (9)$$

This result is close to that reported by Trepanier and Franklin (2010), but markedly different from the results of others (**Figure 14**). Overall, the experimental data from 2D granular column collapse experiments that used the quasi-2D granular flow model shared the same form of relationship between the ratio $(d_\infty - d_0)/d_0$ and the coefficient $a$. These equations can be generalized as follows:

$$\frac{d_\infty - d_0}{d_0} \approx \begin{cases} \lambda_1 a^{\alpha_1} & \text{with } a < a^* \\ \lambda a^\alpha & \text{with } a > a^* \end{cases} \quad (10)$$

where $\lambda_1$ and $\lambda$ are constants that depend on materials used and other parameters fall within the following range:

$$\begin{cases} 0.7 < a^* < 3.0 \\ 0.6 < \alpha_1 < 1.3 \\ 0.59 < \alpha < 0.70 \end{cases} \quad (11)$$

By comparing our experimental data to the general Eq. (10), the following parameters are found for true 2D granular flow:

$$\begin{cases} a^* = 1.5 \\ \alpha_1 = 0.96 \\ \alpha = 0.73 \end{cases} \quad (12)$$

The coefficients $a^*$ and $\alpha_1$ obtained from our experiments fall within the range reported by previous authors who used the quasi-2D model. However, the coefficient $\alpha$ is slightly greater than the values reported by other authors.

**Figure 14** shows a comparison of the experimental results reported by the different research groups. The solid line represents our experimental data using the true 2D granular model, whereas all other dashed lines show results obtained from quasi-2D granular models. To plot the result reported by Balmforth and Kerswell (2005), we used $\lambda = 3.25$ for $a < 1.5$ and $\lambda = 3.8$ for $a \geq 1.5$, which were obtained from our experimental data as shown in Eq. (1). Our experimental data are very close to those reported by Balmforth et al. (2005) for a narrow channel of 1cm width. This suggests that true 2D granular flow is most closely represented by narrower horizontal flow channels when using the quasi-2D approach. However, the data for the true 2D experiment are markedly different from those reported by Balmforth and Kerswell (2005) for a wide flow channel and from other experimental data reported in the literature. These differences can be accounted for by variation of two key reasons across the experiments: 1) the 3D material used and 2) the flow channel width. In all of the quasi-2D experiments discussed above, the 3D materials differ widely in the size and shape of the particles. In addition, various flow channel widths were adopted in these studies. The wider the flow channel, the larger was the deviation from our experiment.

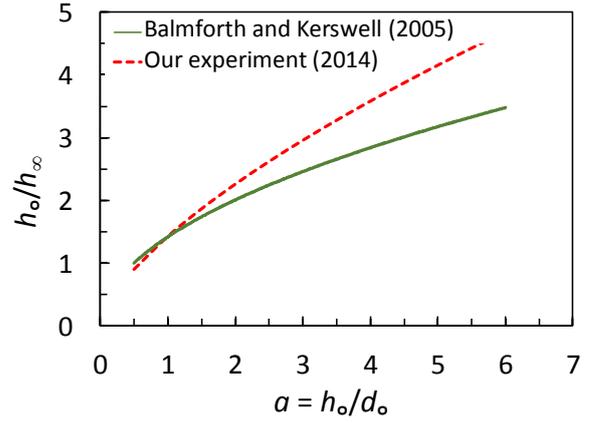

**Fig. 15** Comparison of the relationship between the ratio $h_0/h_\infty$ and the coefficient $a$

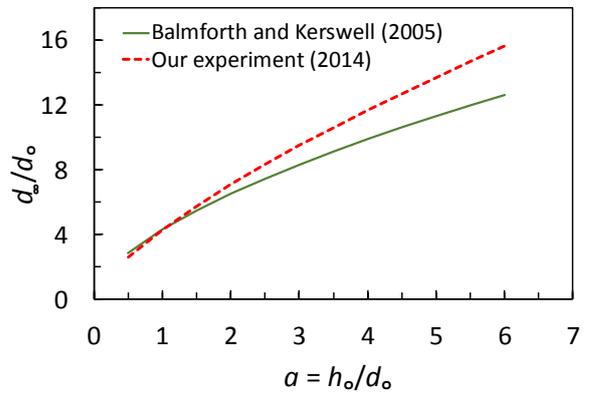

**Fig. 16** Comparison of the relationship between the ratio $d_\infty/d_0$ and the coefficient $a$

The empirical relationship between the ratio $(d_\infty - d_0)/d_0$ and the coefficient $a$ reported by Balmforth and Kerswell (2005) for narrow flow channel fits our 2D experiments data quite well. However, other relationships based on a narrow flow channel, such as that between the ratio $h_0/h_\infty$ or $d_\infty/d_0$ and the coefficient $a$, were markedly different from our data. In particular, Balmforth and Kerswell (2005) reported the following empirical equations:

$$h_0/h_\infty \approx \lambda a^{0.5} \quad (13)$$

$$d_\infty/d_0 \approx \lambda a^{0.55 \pm 0.05} \quad (14)$$

which represent the power-law dependencies of $h_0/h_\infty$ and $d_\infty/d_0$ on the initial aspect ratio $a$. The above empirical equations are consistent with our experimental finding on the true 2D experimental model as expressed by Eqs. (2) and (4), which show that the ratios $h_0/h_\infty$ and $d_\infty/d_0$ are exponential functions of the coefficient $a$. However, the exponential coefficients obtained from true 2D conditions are higher than those obtained from quasi-2D conditions. To further illustrate these differences, Eqs. (13) and (14) are plotted in **Figures 15** and **16**, using values of λ obtained from our experiment (i.e., λ = 1.42 and λ = 4.30, respectively). Our data agree with those of Balmforth and Kerswell (2005), in particular for an initial aspect ratio $a$ of less than 1.5. For a higher $a$ value, our data deviate from their equations. Therefore, we conclude that the quasi-2D experimental model does not fully represent the 2D conditions, even for a very narrow horizontal flow channel. Accordingly, care must be taken when validating 2D numerical models with quasi-2D experimental data.

**Conclusions**

We investigated the failure mechanism of 2D granular flow by using a truly 2D granular flow model (i.e., aluminum rods as the soil model). The results were then compared with experimental data obtained by other authors who used a quasi-2D granular flow model (i.e., a narrow horizontal flow channel with 3D granular soils). Interestingly, our experimental findings were markedly different from those reported by previous authors. We also showed that the quality of the ground contact surface affects the destruction zone and final run-out distance of the granular column after collapsing. In particular, for the same ratio of initial height to initial width of a rectangular column, the final run-out distance of the granular column on a soft ground contact surface was less than that on a hard ground contact surface. A deep-seated failure mechanism was observed in the experiment on the soft ground contact surface.

Finally, our paper provided comprehensive experimental data on the collapsing process of a 2D granular column with full details of the material properties. These data will serve as useful resources to test 2D numerical models to be developed in the near future.

**Acknowledgements**

The first author would like to thank the JSPS (Japan Society for the Promotion of Science) for their financial support through the RONPAKU fellowship (ID number VNM11010).